# Towards a generalized monaural and binaural auditory model for psychoacoustics and speech intelligibility

Thomas Biberger [a)] and Stephan D. Ewert

Medizinische Physik and Cluster of Excellence Hearing4all, Universität Oldenburg, 26111 Oldenburg, Germany.

a) Electronic mail: thomas.biberger@uni-oldenburg.de

Running title: Modeling masking and speech intelligibility

# ABSTRACT

Auditory perception involves cues in the monaural auditory pathways as well as binaural cues based on differences between the ears. So far auditory models have often focused on either monaural or binaural experiments in isolation. Although binaural models typically build upon stages of (existing) monaural models, only a few attempts have been made to extend a monaural model by a binaural stage using a unified decision stage for monaural and binaural cues. In such approaches, a typical prototype of binaural processing has been the classical equalization-cancelation mechanism, which either involves signal-adaptive delays and provides a single channel output or can be implemented with tapped delays providing a high-dimensional multichannel output. This contribution extends the (monaural) generalized envelope power spectrum model by a non-adaptive binaural stage with only a few, fixed output channels. The binaural stage resembles features of physiologically motivated hemispheric binaural processing, as simplified signal processing stages, yielding a 5-channel monaural and binaural matrix feature "decoder" (BMFD). The back end of the existing monaural model is applied to the 5-channel BMFD output and calculates short-time envelope power and power features. The model is evaluated and discussed for a baseline database of monaural and binaural psychoacoustic experiments from the literature.

# I. INTRODUCTION

Auditory perception is typically binaural, involving signals at both ears. Besides enabling localization based on interaural time and intensity differences, interaural disparities can also be exploited to better detect a target stimulus in spatially separated or spatially differently distributed maskers (spatial release from masking, SRM; e.g., [1, 2]) or an antiphasic tone in diotic noise (binaural masking level difference, BMLD; e.g., [3, 4]). Auditory models have been used to explain and analyze monaural and binaural psychoacoustic phenomena (e.g., [5-9]), and as supportive tools offering instrumental assessment of, e.g., speech intelligibility (SI) and audio quality, applicable for development and control of signal processing (e.g., [10-16]). In such applications typically monaural phenomena and perceptive cues involved in, e.g., spectral and temporal masking [17, 18], as well as binaural cues involved in, e.g., sound source location, apparent source width [15], occur in combination [19, 20]. Auditory models as well as psychoacoustic experiments have often focused on either monaural or binaural aspects of perception in isolation, having led to a variety of monaural models (e.g., [ 5, 6, 9, 21, 22, 23, 24, 25, 26]) and binaural models (e.g., [8, 12, 27, 28, 29, 31, 32, 33, 34]). The binaural models typically share "common ground" assumptions of essential monaural preprocessing steps followed by a binaural interaction (BI) stage. In many of these binaural models, the prototype binaural interaction is based on the equalization-cancelation mechanism (EC; [28]) providing a "monaural", single channel output signal after a signal-adaptive binaural noise cancelation. This single channel output either uses the optimal internal delay to compensate for external interaural delays in connection with an optimal level compensation (equalization) to cancel undesired noise, comparable to an adaptive binaural (or bilateral) beamformer (for an overview, see [35]), or simply selects the better ear (referred to as "better-ear glimpsing" if applied in time-frequency frames, see [1]). Thus, the EC mechanism can be easily applied as binaural front end to an existing monaural model (for speech intelligibility see, e.g., [12, 13, 14, 36, 37]). Providing a monaural or diotic input, reverts such models to monaural ones, although they are typically applied to binaural (dichotic) stimuli. Focusing on a large variety of basic binaural psychoacoustic experiments, Breebaart et al. [8, 38, 39] combined a number of internal delays and interaural gains in a matrix of (excitatory-inhibitory) cancelation elements. By this, a signal-adaptive mechanism to equalize prior to cancelation as in the EC approach is avoided, however, a signal-adaptive template mechanism is required to "select" optimal matrix elements by applying weights in the form of a template for a given psychoacoustic experiment. Both the monaural front end and the template-matching procedure used in the Breebaart model have been taken from the (monaural) perception model of Dau et al. [5, 6].

The above mentioned models show successful concepts for combining monaural and binaural model stages in a combined model, however, they have been either explicitly applied to binaural psychoacoustics or speech intelligibility whereas their front and back ends without binaural stage have been explicitly applied to the respective monaural experiments. Moreover, the models require a signal-adaptive mechanism in the EC stage and a selection from 3 output channels (EC approach: Left, EC output, right) or a signal-adaptive template to extract information from the high-dimensional matrix of delay-gain elements.

The question arises whether a simpler, non-adaptive approach is sufficient to model binaural interaction. For speech intelligibility in symmetrically placed interferers, e.g., [2] found that a simple addition of the left and right input channel can explain a large part of the observed spatial release from masking (SRM). Such a simplistic binaural interaction has also been suggested by [40] as midline spatial channel in the human auditory cortex. Additionally, the existence of delay lines as utilized in the EC and Breebaart approach has been questioned in mammals (for a review see [41]) and physiologic studies (e.g., [42, 43]) suggest a simpler hemispheric model without delay lines to account for binaural interaction, involving fixed phase delays and excitation as well as inhibition from the contralateral ear. Regarding the development of effective auditory signal processing models, such a fixed binaural interaction could be beneficial for applications where computational efficiency is important. Moreover, it appears desirable to evaluate the same model both in monaural and binaural experiments as well as in basic psychoacoustic tasks and speech intelligibility. The advantage of such a unified modelling approach (see, e.g., [9, 26] for monaural models) is the applicability of the model to a wide variety of stimuli as well as the potential of the model to directly link performance and cues in basic psychoacoustic tasks, such as detection and discrimination thresholds, to higher level processes involved in speech intelligibility. In the long run, such a link might help to understand and disentangle peripheral and central deficits in hearing impaired and elderly persons (e.g., [44 - 48]) and in the context of model-driven stimulus design for psychoacoustics and physiology (e.g., [49]).

Here we suggest and examine a combined monaural and binaural model in a variety of „benchmark" psychoacoustic and speech intelligibility experiments. The combined approach uses the monaural front end and back end of the generalized power spectrum model (GPSM; [26]) which has been successfully applied to monaural psychoacoustics, speech intelligibility and audio quality ([9, 16, 19, 20, 26]). A binaural processing stage with five fixed (non-adaptive) output channels is suggested prior to the model back end, referred to as binaural matrix feature decoder (BMFD). The output comprises the left (L) and right (R) channels, the L+R channel and the L-R and R-L channels, incorporating a fixed phase delay and gain. L and R enable better ear glimpsing in connection with a selection of time-frequency frames across the BMFD output channels in the back end (better ear channels). The three other channels realize a binaural interaction: L+R represents a midline channel, enhancing coherent (frontal) signals at both ears. The L-R and R-L channels effectively mimic the outputs expected in hemispheric models of binaural interaction in a highly simplified manner. These channels are comparable to two elements in the delay-gain matrix of the Breebaart model, or to two according parameter choices in the EC approach. The ability of the suggested model to account for the monaural and binaural data and the relevance of the five BMFD output channels are assessed in the following.

## II. Model description

The front end of the proposed GPSM with BMFD extension calculates short-time power and envelope power features for each of two better-ear (BE) channels (L: $BE_L$, R: $BE_R$) and the three binaural interaction (BI) channels (L-R: $BI_L$, L+R: $BI_C$, R-L: $BI_R$), comprising the binaural matrix feature decoder. Signal-to-noise ratios based on these features are assessed by a task-dependent decision stage (psychoacoustics or speech intelligibility) in the model back end. The model processes two input stimuli, the target-plus-masker (signal) and masker alone (noise).

### A. Monaural processing stages

The peripheral processing, feature extraction and decision stage of the GPSM with BMFD extension, illustrated in Figure 1 are similar to that of the monaural mr-GPSM proposed in [26]. In the following, the processing stages related to the envelope power pathway are only roughly described here, and for a more comprehensive description the reader is referred to [9, 26].

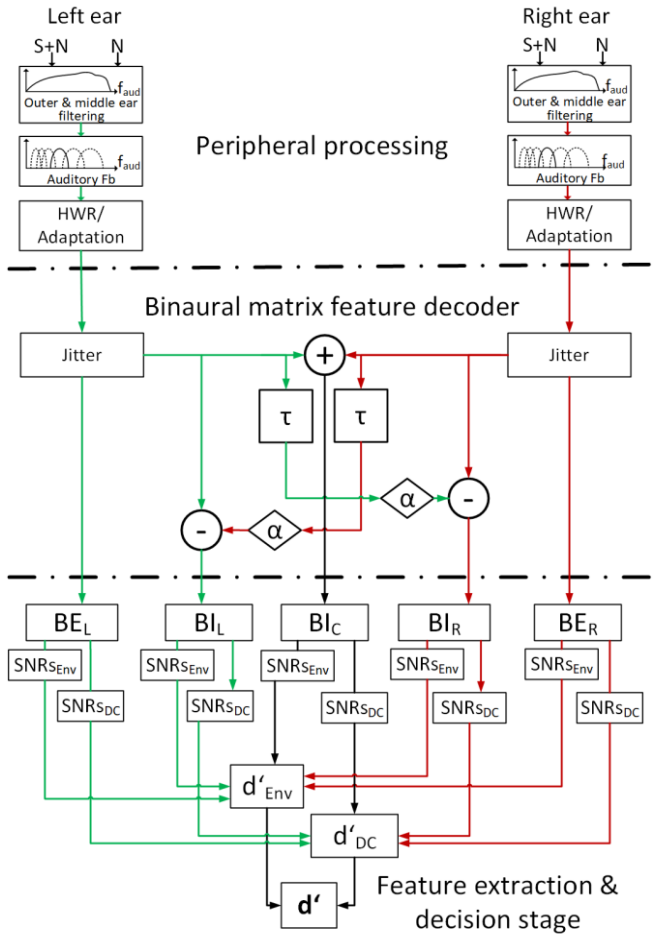

Figure 1: Block diagram of the GPSM with BMFD extension. After peripheral processing, the left and right ear signals are binaurally processed by using the BMFD that provides two better-ear channels $BE_L$ and $BE_R$ and three binaural interaction channels $BI_L$, $BI_C$, $BI_R$. For each of the five BMFD outputs, envelope power and power SNRs are calculated in short-time frames and then combined across the five channels of the BMFD and across auditory and modulation channels, resulting in a sensitivity index $d'_{env}$ based on envelope power SNRs and $d'_{DC}$ based on power SNRs. The final combined $d'$ is then compared to a threshold criterion that assumes that a signal is detected if $d' > (0.5)^{1/2}$.

The initial *Outer & middle ear filtering* stage (see Figure 1) weights the input signal with the hearing threshold in quiet [50], followed by the *Auditory Fb*, reflecting basilar membrane filtering by applying a fourth-order Gammatone filterbank with bandwidth equal to the equivalent rectangular bandwidth of the auditory filter (ERB$_N$; [51]) and third octave spacing from 63 to 12500 Hz. In contrast to Hilbert envelope extraction in [26], each auditory channel is half-wave rectified to simulate that inner hair cells primarily respond only to one direction of deflection. The half-wave rectified signals are divided by an integrator with time constant of 2 ms, realized as a first-order low pass filter with cut-off frequency of 500 Hz, to simulate effects of neural adaptation of the auditory system in a simple feed-forward manner.

**B.   Binaural processing stages**

The adapted signals from the monaural processing of the left and right ear serve as input for the binaural processor. First, amplitude and phase jitter are applied independently for each auditory channel to the input signals, to limit the performance of the BI. Amplitude and time jitters are generated as zero-mean Gaussian processes with a standard deviation of $\sigma_\epsilon = 0.25$ and $\sigma_\delta = 105$ µs, as suggested by [28] and also applied by [36] and [37]. Based on the jittered signals three BI channels BI$_L$, BI$_C$, and BI$_R$ are calculated according to Eq. 1-3:

$$\mathrm{BI_L}(p,t) = L(p,t) - \alpha\, R\big(p, t - \tau(p)\big) \tag{1}$$

$$\mathrm{BI_C}(p,t) = \sqrt{L(p,t)\, R(p,t)} \tag{2}$$

$$\mathrm{BI_R}(p,t) = R(p,t) - \alpha\, L\big(p, t - \tau(p)\big). \tag{3}$$

BI$_L$ results from subtracting the time delayed and amplified right ear channel $\alpha \cdot R\big(p, t - \tau(p)\big)$ from the left ear channel $L(p,t)$ in each auditory channel $p$. BI$_R$ is calculated vice versa to BI$_L$. Based on physiologic findings and preliminary tests, a frequency-dependent delay τ equal to a phase shift of π/4 was chosen, resulting in longer delays for lower frequencies. The amplification factor α equals 3 (see discussion for further details). BI$_C$ accounts for the effect of adding the left and right ear signals prior to auditory processing. Taking the half-wave rectified signal representation into account, this is achieved by the square root of the product $L(p,t)$ and $R(p,t)$, making BI$_C$ a midline channel most sensitive to sound images spatially placed in the median plane. In addition to the three BI channels, the (monaural) left and right channel $L(p,t)$ and $R(p,t)$ are passed unaltered as output of the five channel BMFD stage. They can be used for better-ear glimpsing in the following feature extraction stage (referred to as BE$_L$, BE$_R$).

### C. Power and envelope power feature extraction stage

A first-order low-pass filter with cut-off frequency of 150 Hz [7, 52] is applied to the five output channels of the BMFD. The consecutive processing stages in each of the five BMFD channels are separated into two independent pathways where envelope power SNRs (EPSM; left-hand side of Figure 1), and power SNRs (PSM; right-hand side of Figure 1) are calculated. Indices for the BMFD channels are omitted for clarity in the following equations.

In the PSM path, the intensity (DC-power) features $P_{DC,j}(p)$ are calculated in short-time windows $j$ by taking the squared mean of the Hilbert envelope within each auditory channel $p$

$$P_{DC,j}(p) = \frac{\left[\overline{E}_j(p)\right]^2}{2}. \tag{4}$$

The duration of the windows depends on the center frequency of the auditory channel, where the lowest center frequency of 63 Hz corresponds to window length of 45 ms and the highest center frequency provides a window length of 8 ms. As proposed by Rhebergen and Versfeld [11] values for the window duration were taken from [53] and multiplied by 2.5. Intensities $P_{DC,j}(p)$ falling below the hearing threshold are set to 1e-10. Then the $\text{SNR}_{DC,j}(p)$ is calculated between target-plus-masker intensities $P_{DC,\text{targ+mask},j}(p)$ and the masker intensities $P_{DC,\text{mask},j}(p)$ according to

$$\text{SNR}_{DC,j}(p) = \frac{P_{DC,\text{targ+mask},j}(p) - P_{DC,\text{mask},j}(p)}{P_{DC,\text{mask},j}(p)}. \tag{5}$$

For speech intelligibility predictions, optionally a band importance function (BIF) as used in the ESII, is multiplicatively applied to the intensity $\text{SNR}_{DC}(p)$. Note that the here applied BIF is normalized by its highest value and thus the $\text{SNR}_{DC}$ within this auditory channel remains unaffected from the (normalized) BIF, while all other channels become attenuated. In the EPSM path, the envelopes are initially processed by a modulation filterbank consisting of bandpass filters ranging from 2 to 256 Hz with a Q-value of 1 and a third-order low-pass filter with cut-off frequency of 1 Hz. Hereby, based on [54], only modulation filter center frequencies up to one fourth of the corresponding auditory channel center frequency are considered. Then the AC-coupled envelope power $P_{\text{env},j}(p,n)$ is calculated for each auditory channel $p$, modulation channel $n$, and time window $i$, as it was proposed in [25], by applying a lower limit of -27 dB for the envelope power, reflecting the limitation in human sensitivity to amplitude modulation (AM) [22, 52]. The envelope power based signal-to-noise ratio $\text{SNR}_{\text{env},i}(p,n)$ between the target-plus-masker and masker envelope power is calculated according to [25] and then a logarithmic weighting of envelope power SNRs is applied for auditory channels with intensity levels of the target-plus-masker stimuli below 35 dB, while envelope power SNRs above that level are unaffected from weighting.

Taken together, the output of the model front end consists of intensity weighted envelope power SNRs, $\text{SNR}_{\text{envW},i}(p,n)$, and power SNRs, $\text{SNR}_{\text{DC},j}(p)$, for each of the five BMFD output channels.

### D. Decision stage

The envelope power and power based SNRs are subjected to a task-specific decision stage for predicting psychoacoustic detection or discrimination thresholds and SI data.

#### 1. *Psychoacoustics*

In the first step, $\text{SNR}_{\text{envW},i}(p,n)$ in each of the five front end output channels are combined by taking the largest value for each time frame within each auditory and modulation channel resulting in $\text{SNR}_{\text{envWC},i}(p,n)$. $\text{SNR}_{\text{envWC},i}(p,n)$ is then averaged across temporal segments *i* per modulation filter, resulting in a two-dimensional representation of envelope power $\text{SNR}_{\text{env}}(p,n)$. The same procedure is applied to combine $\text{SNR}_{\text{DC},j}(p)$ across the five channels resulting in the $\text{SNR}_{\text{DCW},j}(p)$ which is then is averaged across temporal segments *j*, resulting in a 1-dimensional representation of power SNRs over auditory channels denoted as $\text{SNR}_{\text{DC}}(p)$

Finally, the envelope power and power SNRs [$\text{SNR}_{\text{env}}(p,n)$, $\text{SNR}_{\text{DC}}(p)$] are combined in the same manner as proposed in [26]:

$$\text{SNR} = \max\left[\beta \cdot \left(\sum_{p=1}^{M}\sum_{n=1}^{N}\text{SNR}^2_{\text{env}(p,n)}\right)^{\frac{1}{2}}, \gamma \cdot \left(\sum_{p=1}^{M}\text{SNR}^2_{\text{DC}(p)}\right)^{\frac{1}{2}}\right] \quad (6)$$

At first envelope power and power SNRs are combined across auditory and modulation channels (in case of envelope power) and auditory channels [inner brackets in Eq. 6] and then multiplied with empirical determined correction factors $\beta = 0.21$ and $\gamma = 0.45$. Both correction factors are identical to those proposed in [9, 26] and are used due to violation of the assumption of independent observations in the auditory and modulation channels, because of using overlapping bandpass filter. Finally, the domain (envelope or power), providing the highest SNR-value is chosen.

As in [9, 26] the decision criterion used in this study is based on [7] assuming that a signal is detected if the SNR > -6 dB (equivalent to a power ratio of 0.25), which can, according to [55] also be expressed as sensitivity index $d' = (2 \cdot \text{SNR})^{1/2} \approx (0.5)^{1/2}$.

### *2.   Speech intelligibility*

The overall SNR is obtained by applying the same procedure as described for psychoacoustic predictions. The overall SNR is converted to the sensitivity index $d'$ by using equation (6) from [25] and finally transformed into percent correct responses.

### E.   Model configurations

All model versions with binaural extension tested in this study had the same settings as the monaural GPSM-versions in [9, 26]: For psychoacoustic experiments, auditory filters had a third-octave spacing ranging from 63 to 12500 Hz, while auditory filters range from 63 to 8000 Hz for SI experiments. For SI predictions, the band-importance weighting, as it was proposed by Table 3 of [56] was exclusively applied to the power SNRs. Each of the models used exactly the same set of parameters for all experiments.

## III. Psychoacoustic evaluation

### A. Monaural experiments

In this study the same set of headphone-based monaural psychoacoustic experiments were applied for model evaluation as in [9, 26]. Thus, these experiments are only briefly explained in the following. For more detailed information the reader is referred to [9] or the respective original publications.

Experiment 1 (Intensity discrimination and hearing thresholds). Just noticeable intensity level differences (JNDs) as a function of the reference level (20, 30, 40, 50, 60, 70 dB) were measured for a 1-kHz pure-tone (in quiet) and broadband noise ranged from 0.1 to 8 kHz [57]. The target interval contained an increased level $L_t = L_0 + \Delta L$ where $L_0$ corresponds to the reference level and $\Delta L$ corresponds to the JND, which can be rewritten in terms of intensities as $\Delta L = 10 \log_{10} \frac{I_t}{I_o} = \frac{\Delta I + I_0}{I_o}$. Hearing thresholds ranging from 50 Hz to 10 kHz were taken from [50].

In Experiment 2 (Spectral masking with narrow-band and pure-tone maskers) the masking patterns for four different signal-masker combinations of noise-in-tone (NT), noise-in-noise (NN), tone-in-tone (TT) and tone-in-noise (TN) originated from [58]. The noise corresponds to a Gaussian noise with a bandwidth of 80 Hz, while the tone refers to a sinusoidal stimulus. The masker had a fixed center frequency at 1 kHz, while the signal had frequencies of 0.25, 0.5, 0.75, 0.9, 1.0, 1.1, 1.25, 1.5, 2, 3, and 4 kHz. All signal-masker combinations, with exception of the TT condition, where each stimulus had a fixed phase of 90°, had random phases. Data for the masker levels of 45 and 85 dB are considered here.

Experiment 3 (Tone in noise masker) was taken from [24] and reflects detection thresholds of a 2-kHz pure tone signal in the presence of a band limited (0.02 to 5 kHz) Gaussian noise masker for signal durations from 5 to 200 ms. The masker had a duration of 500 ms and the signal was temporally centered in the masker. The presentation level of the masker was 65 dB SPL.

Experiment 4 (AM-depth discrimination) is based on the study from [59] where AM-depth discrimination function for a 16 Hz sinusoidal AM with respect to fixed reference AM-depths was measured for sinusoidally modulated broadband noise (1.952-4 kHz) and pure-tone carriers (4 kHz) at an overall presentation level of 65 dB SPL. The AM depth of the (standard) reference signal $m_s$ ranged, in 5-dB steps, from -28 to -3 dB. The increased AM depth of the target signal is given by $m_c = m_s\sqrt{1 + m_{inc}}$. Within the measurement the fractional increment $m_{inc} = (m_c^2 - m_s^2)/m_s^2$ was varied in dB ($10 \log m_{inc}$).

In Experiment 5 (AM detection) temporal modulation transfer functions (TMTF) for three narrow band noise carriers of 3, 31, and 314 Hz [5] and broadband noise carriers [22] were considered. The narrow band noise carriers were centered at 5 kHz and a sinusoidal AM of 3, 5, 10, 20, 30, 50, and 100 Hz was used. The narrow band carrier level was 65 dB SPL and the stimuli were adjusted to have equal power after AM. The broadband noise carriers ranged from 0.001 to 6 kHz and a sinusoidal AM of 4, 8, 16, 32, 64, 128, 256, 512, and 1024 Hz was applied. The level of the broadband carriers was 77 dB SPL.

Experiment 7 (Amplitude modulation masking) was taken from [9] and measured AM masking and detection thresholds for a target sinusoidal amplitude modulation (SAM) in the presence of a sinusoidal or squarewave masker modulation. The effect of varying the carrier type (broadband and pure-tone carriers), masker waveform (sinusoidal or squarewave), and modulation rate of the target (4 and 16 Hz) and masker (16 and 64 Hz) were examined in four different stimulus configurations which can be seen in Table 1 of [9].

**B.    Binaural experiments**

Six binaural headphone experiments from literature were used for the model evaluation. The maskers used in the binaural experiments had a duration of 400 ms unless otherwise stated. In several binaural experiments target and masker signals comprise interaural manipulations indicated by subscripts: The subscript 0 indicates no interaural phase shift (in phase), the subscript $\pi$ indicates an interaural phase shift of $\pi$ (out of phase), and the subscript m indicates that the corresponding signal was presented monaurally. Accordingly, a $N_0S_\pi$ stimulus indicates that the noise signal $N_0$ is interaurally in phase, while the target signal $S_\pi$ is interaurally out of phase. The experiments are only briefly described in the following and the reader is referred to [38, 39] for experiment 1-5 or the original literature for further details.

Experiment 1 (ITD discrimination) is based on the ITD experiments from [60, 61], where discrimination threshold for ITDs were measured for pure tone stimuli at various frequencies. The reference stimuli were presented dioticaly at a level of 65 dB SPL, while the target stimuli were presented at the same level but had an ITD. The tested frequencies ranged from 90 to 1500 Hz.

Experiment 2 (IID discrimination) is based on the IID experiments from [62, 63], where thresholds for IID were measured for pure tones at various frequencies ranging from 62.5 to 4000 Hz. The reference stimuli were presented dioticaly at a level of 65 dB SPL. The target stimuli had an IID, resulting in an overall level of (65+IID/2) dB SPL for the left channel and (65 – IID/2) dB SPL for the right channel.

Experiment 3 (Frequency and interaural phase relationships in wideband conditions) is based on experiments of [3, 4, 64, 65], where thresholds of the four binaural conditions $N_0S_\pi$, $N_\pi S_0$, $N_0S_m$, and $N_\pi S_m$, were measured as a function of the frequency of the pure tone signal (125, 250, 500, 1000, 2000, and 4000 Hz). The masker was a low-pass noise with a cutoff frequency of 8 kHz and a spectral level of 40 dB/Hz.

Experiment 4 ($N_0S_\pi$ depending on signal duration) is based on experiments of [66-69], where $N_0S_\pi$ detection thresholds were measured as a function of the target signal ($S_\pi$) duration. The masker signal ($N_0$) was a 500-ms wideband noise with a spectral density of 36.2 dB/Hz. The target signal was a pure tone of either 500 Hz or 4 kHz with signal durations ranging from 2 to 256 ms.

Experiment 5 (Temporal phase transition) is based on the experiments of Kollmeier and Gilky [70] where $N_0N_\pi S_\pi$, $N_\pi N_0 S_\pi$, $N_\pi N_{\pi,-15dB} S_\pi$, $N_{\pi,-15dB} N_\pi S_\pi$, thresholds were measured as a function of the temporal position of the target signal ($S_\pi$) relative to the masker-phase transition ($N_\pi N_0$ or $N_0 N_\pi$) to estimate the temporal resolution of the binaural auditory system. The broadband noise maskers with a duration of 750 ms were bandpass filtered from 100 to 2000 Hz and had a spectral level of 40 dB/Hz. The $N_0N_\pi$ masker started with an interaural phase of $N_0$ that switched to $N_\pi$ after 375 ms. Accordingly, $N_\pi N_0$ started with a 375 ms interaurally out of phase segment followed by a 375 ms in phase segment. The interaurally out of phase masker $N_\pi N_{\pi,-15dB}$ was attenuated by 15 dB 375 ms after its onset. The interaurally out of phase masker $N_{\pi,-15dB} N_\pi$ was amplified by 15 dB 375 ms after its onset. $S_\pi$ was an interaurally out of phase pure tone of 500 Hz with a duration of 20 ms. The masked threshold was measured as a function of the delay time between the transition of the noise segments and the signal offset.

Experiment 6 (Time-intensity-trading) is based on experiments of Hafter and Carrier [71], where $d'$ was measured for several combinations of fixed ITDs (0, +10, +20, +30, and +40 µs; positive sign indicates left ear leading) and varying IIDs (ranging from 0 to -3 dB; negative sign indicates right ear more intense) to examine to which extent time differences can be traded against level differences. The reference signal was a diotic pure tone of 500 Hz (centered sound image). The test signal had a ITD promoting lateralization to the left side, and a IID promoting lateralization to the right side. The lowest $d'$ measured for a certain IID at a fixed ITD indicates that the test signal was most similar to a centered image.

## C. Results and discussion

Predictions from three model versions were compared to disentangle the contribution of the binaural interaction ($BI_L$, $BI_C$, $BI_L$) and better-ear ($BE_L$, $BE_R$) BMFD channels. Model predictions based on all five channels are abbreviated as BMFD and represented by open circles. Model predictions based on the three binaural interaction channels are abbreviated as $BI_{L,C,R}$ (open squares), while predictions based on only the left and right BI channel are abbreviated as $BI_{L,R}$ (open diamonds).

### 1. *Monaural Experiments*

The upper part of Table 1 reports root-mean squared errors (RMSEs) and the coefficient of determination ($R^2$) between experimental data and predictions based on BMFD, $BI_{L,R}$, and the monaural mr-GPSM [26]. For the monaural experiments stimuli were only provided to the left-ear input channel of the BMFD and the right-ear input channel was set to zero. As obvious from the RMSE- and $R^2$-values, BMFD predictions largely agree with those from the monaural mr-GPSM. Given the similarity of both models for the monaural data, detailed figures to compare the subjective and predicted data are not shown here. The similarity is expected as the BMFD has only a few modifications which potentially influence monaural prediction performance. As shown in Table 1, prediction performance was not degraded when only $BI_L$ and $BI_R$ ($BI_{L,R}$) were used instead of all five BMFD outputs. This result was also expected, because when the right input channel is set to zero, $BI_L$ only depends on the left ear channel, and in such monaural conditions $BI_L$ is equal to $BE_L$. Accordingly, reducing the number of output channels of the BMFD would be sufficient to capture important monaural psychoacoustic effects, but may not sufficient to account for all the binaural aspects assumed to be important to explain a variety of data from binaural psychoacoustic and SI experiments.

To summarize, for monaural experiments tested in this study the GPSM with binaural BMFD extension largely maintains the prediction performance of the monaural mr-GPSM.

## *2.    Binaural Experiments*

In Figures 2 – 6, subjective and predicted data for the binaural experiments are represented by closed and open symbols, respectively. The lower part of Table 1 reports root-mean square errors (RMSE) and the coefficient of determination ($R^2$) between experimental data and predictions based on BMFD, $BI_{L,C,R}$, and $BI_{L,R}$.

As illustrated in the upper panel of Figure 2, data of [60, 61] showed that ITD thresholds decreases with increasing target tone frequency, where the smallest ITD threshold of about 0.012 ms was found at 1 kHz. These decreasing threshold ITDs represent a more or less constant IPD of about 0.05 rad (~ 3°). For frequencies above 1 kHz, measured ITD thresholds increase, which is due to a reduced phase-locking ability of the IHCs for higher frequencies. For all three model versions, predicted ITD thresholds are higher than observed in the data, particularly at low frequencies. Here a nearly constant IPD of about 0.07 – 0.08 rad (~ 4°-5°) was predicted, which is higher than the nearly constant IPD of about 3° in the data. In agreement with the data, predicted ITD thresholds decrease with increasing frequency reaching a plateau at 500 Hz and above. At about 700 Hz, all three models predicted the lowest ITD threshold of about 0.023 μs. For frequencies above 900 Hz $BI_{L,R}$ predictions showed increased ITD thresholds, while predictions based on $BI_{L,C,R}$ and BMFD showed increased thresholds up to about 1200 Hz followed by slightly decreased threshold up to 1500 Hz. For all three model versions ITD thresholds slightly decrease for frequencies above 1.5 kHz.

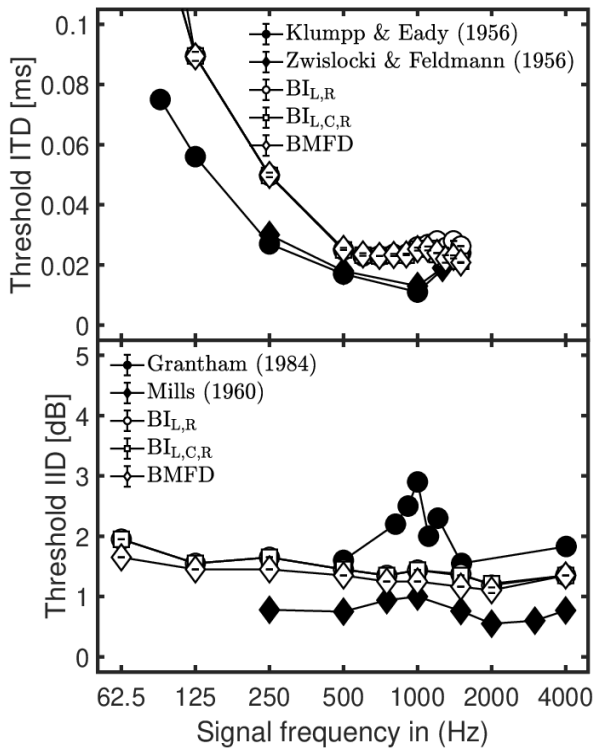

Figure 2: Empirical data (filled symbols) and model predictions (open symbols) for ITD thresholds in ms (upper panel) and IID thresholds in dB (lower panel).

The lower panel of Figure 2 shows measured IID thresholds adopted from the studies of [62, 63]. Across frequencies ranging from 250 Hz to 4 kHz, Mills [62] measured rather similar IID thresholds (average threshold of about 0.8 dB), where the maximum of about 1 dB was reached at 1 kHz. Grantham [63] observed overall about 1.3 dB higher IID thresholds with substantially increased thresholds around 1 kHz. Predicted IID thresholds for the three model versions slightly decreased from about 2 dB at 62.5 Hz to about 1.1 dB at 2 kHz, and increased again for higher frequencies. The predicted IID pattern agrees well with the average of both data sets. Predicted thresholds for $BI_{L,R}$, and $BI_{L,C,R}$ between frequencies from 62.5 Hz to 2 kHz are on average 0.2 dB higher than those from BMFD.

The upper four panels of Figure 3 show measured $N_0S_m$, $N_\pi S_m$ $N_0S_\pi$, $N_\pi S_0$, thresholds adopted from the studies of [3, 4, 64, 65]. All threshold patterns show a V shape with a minimum at 250 Hz. For the monaural target ($S_m$) thresholds are lower for $N_0S_m$ than for $N_\pi S_m$, while for the binaural target ($S_\pi$ or $S_0$) thresholds are lower for $N_0S_\pi$ than for $N_\pi S_0$. The resulting threshold differences of $N_\pi S_m - N_0 S_m$ and $N_\pi S_0 - N_0 S_\pi$ are shown in both lower panels of Figure 3. The largest differences, up to about 9.5 dB, occur for signal frequencies below 500 Hz. $BI_{L,R}$ predictions (open circles) show a similar overall pattern to the data, and accordingly the predicted $N_\pi S_m - N_0 S_m$ and $N_\pi S_0 - N_0 S_\pi$ patterns largely agree with data. For $N_\pi S_m$ and $N_\pi S_0$, both middle panels in Figure 3 show larger deviations between the data and the $BI_{L,C,R}$ and BMFD predictions at 250 Hz and 500 Hz. This deviation is based on the contribution of the $BI_C$ channel that overestimates human performance for the $N_\pi S_m$ and $N_\pi S_0$ conditions. Accordingly large deviations between data and predictions are observed in the difference patterns in the lower two panels for $BI_{L,C,R}$ and BMFD at 250 Hz.

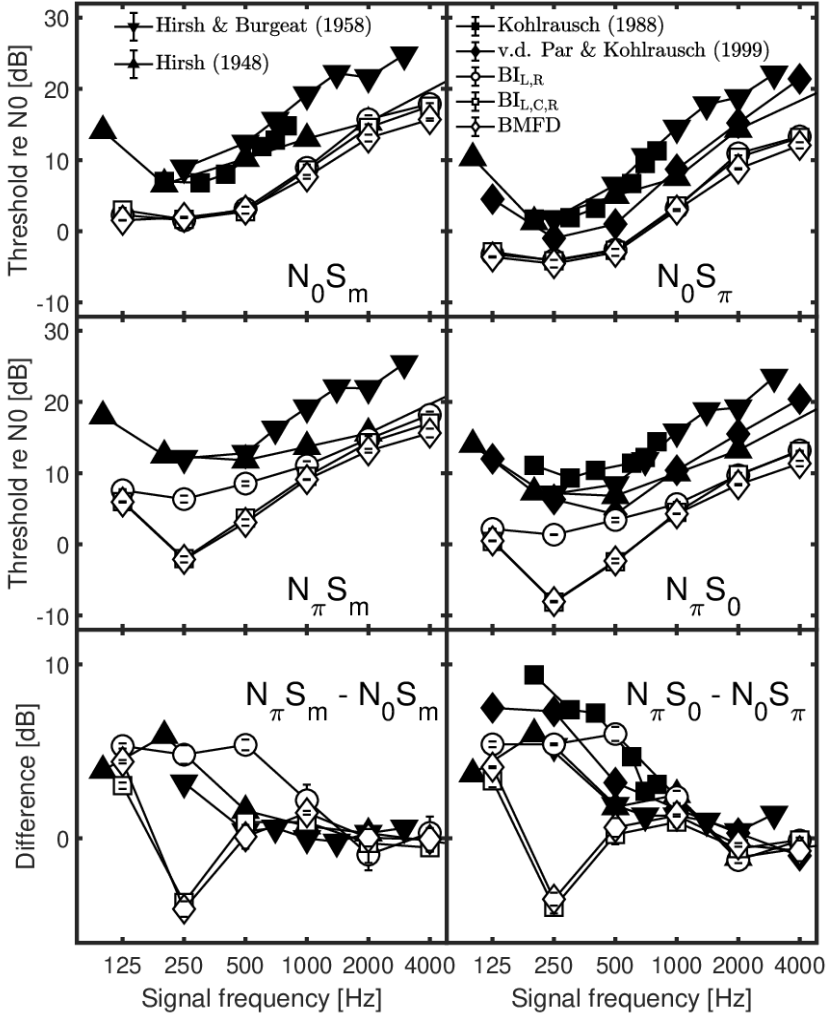

Figure 3: Empirical data (filled symbols) and model predictions (open symbols) for masked thresholds for wideband $N_0S_m$ (upper-left panel), $N_0S_\pi$ (upper-right panel), $N_\pi S_m$ (middle-left panel), and $N_\pi S_0$ (middle-right panel) conditions as a function of the frequency of the signal. Differences in thresholds between the $N_\pi S_m$ and $N_0S_m$ are shown in the lower-left panel, while the lower-right panel represents differences in threshold between $N_\pi S_0$ and $N_0S_\pi$.

Measured $N_0S_\pi$ thresholds as a function of signal duration adopted from [66-69] are shown in Figure 4. For the target signal with frequency of 500 Hz, thresholds decrease with a slope of about 4.5 dB per duration doubling, while for longer signal durations a slope of about 1.5 dB per duration doubling is observed. For the 4 kHz target signal, the data shows a slope of about 3 dB per duration doubling. For all three model versions, nearly identical thresholds were observed with on average higher thresholds than observed in the data. For both signal frequencies predicted thresholds decreased with about 3 dB per doubling of the signal duration, as the signal's energy increases by 3 dB per duration doubling. Such increase in signal duration means that more short-time frames of the model provide an SNR-advantage, that effectively lowers the threshold.

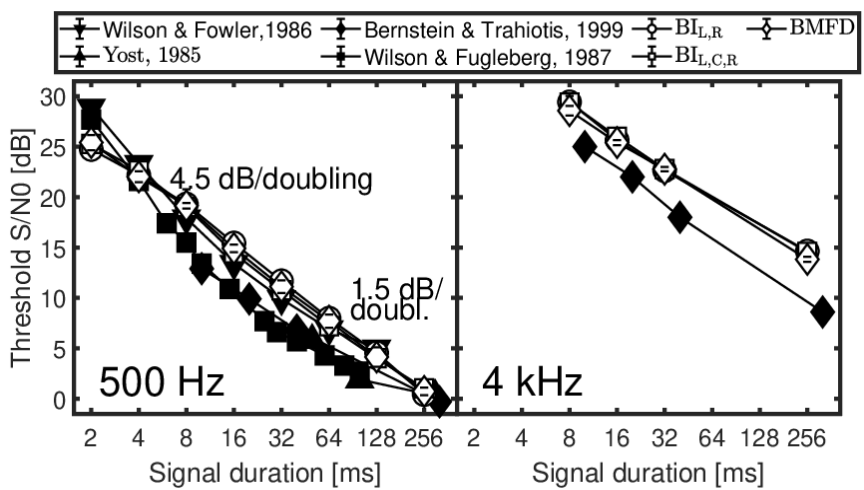

Figure 4: Empirical data (filled symbols) and model predictions (open symbols) for $N_0S_\pi$ thresholds as a function of the signal duration. Data and predictions are shown for signal frequencies of 500 Hz (left panel) and 4 kHz (right panel).

In Figure 5, masked thresholds from four subjects measured by Kollmeier and Gilky [70] are shown. In $N_0N_\pi S_\pi$ and $N_\pi N_0 S_\pi$ conditions lower thresholds (large BMLD) were measured for target signals ($S_\pi$) in the interaurally in phase masker segments ($N_0$) than for $S_\pi$ in interaurally out of phase masker segments ($N_\pi$). Similarly for the corresponding "monaural" $N_\pi N_{\pi,-15dB} S_\pi$ and $N_{\pi,-15dB} N_\pi S_\pi$ conditions, $S_\pi$ in attenuated $N_\pi$ segments resulted in lower thresholds compared to $S_\pi$ in not attenuated $N_\pi$ segments. While a gradual release from masking was observed when shifting $S_\pi$ from the $N_\pi$ segment into the $N_0$ segment (upper-left panel), a very steep release from masking was observed for the corresponding "monaural" $N_\pi N_{\pi,-15dB} S_\pi$ condition (lower-left-panel). A similar behavior was found for the $N_0 N_\pi S_\pi$ and the $N_{\pi,-15dB} N_\pi S_\pi$ conditions. Similar predicted masked thresholds are observed for the three model versions and the predicted steepness of the transition is the same for all four conditions. The predicted BMLD in $N_\pi N_0 S_\pi$ (upper-left panel) and the predicted masking effect in $N_0 N_\pi S_\pi$ (upper-right panel) are somewhat smaller than observed in data. Overall, the predictions largely agree to experimental data, which is also indicated by reasonable RMSE and $R^2$ values of about 2.7 dB and 0.8, respectively.

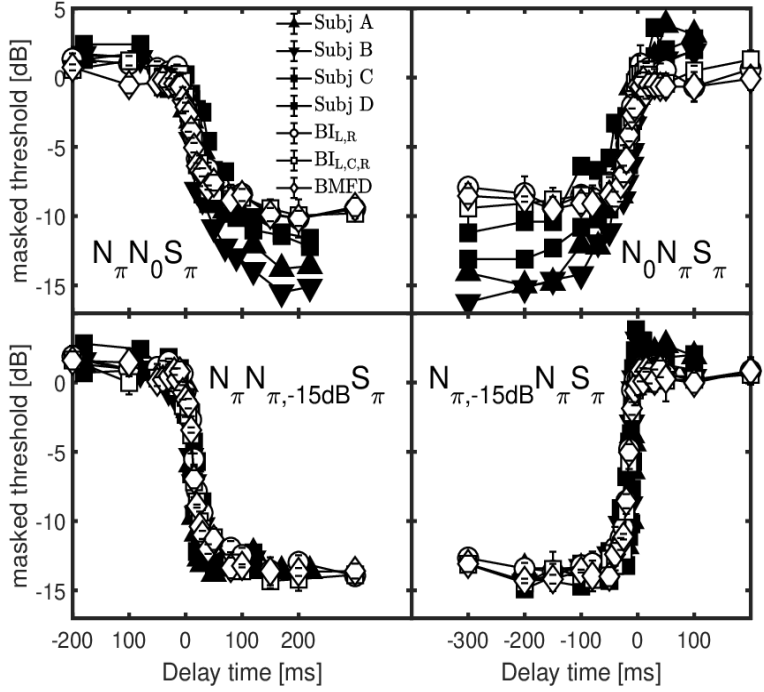

Figure 5: Empirical data (filled symbols) and model predictions (open symbols) for $N_\pi N_0 S_\pi$ (upper-left panel) and $N_\pi N_0 S_\pi$ (upper-right panel) thresholds as a function of the temporal position of the signal center relative to the masker-phase transition. Monaural thresholds for $N_\pi N_{\pi,-15dB} S_\pi$ and $N_{\pi,-15dB} N_\pi S_\pi$ are shown in the lower-left and lower-right panels. Filled symbols represent four subjects measured by Kollmeier and Gilky [70].

The upper and lower panel of Figure 6 show measured d's from the time-intensity-trading experiment of subject S1 and S4 from Hafter and Carrier [71], respectively (see their Figure 1). For clarity only these two subjects with the largest difference in performance are shown in different panels. Likewise, the model predictions for the BI channels and all five channels are split to the two panels for better visibility. Both subjects show that for increasing ITD of 0, 10, 20, 30, and 40 µs a larger opposing ILD was required for "trading" yielding the lowest sensitivity index d' for discrimination of the trading stimulus from the diotic reference signal. It is obvious that the model based on only the BI channels (upper panel of Figure 6) can only mimic the general pattern while there are large differences in the sensitivity and the ILD required for trading as a function of ITD. Moreover, the model with all five BMFD output channels (lower panel of Figure 6) shows even larger deviations to the data and fails to predict a clear dependency of ILD on ITD. Overall the model is closer to the performance of subject S4 than to S1.

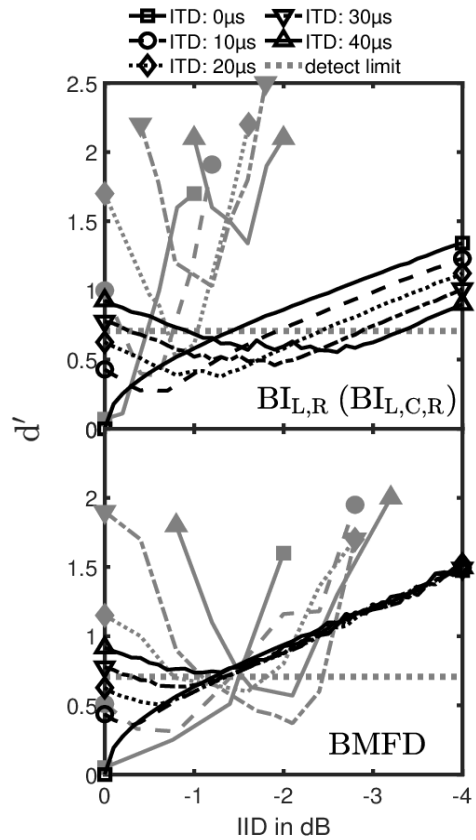

Figure 6: Empirical data (grey lines, closed symbols) and model predictions (black lines, open symbols) for the time-intensity trading experiment of Hafter and Carrier [71] with different ITDs of 0, 10, 20, 30, and 40 µs. The ordinate represents d', while the abscissa represents the IID in dB. Since $BI_{L,R}$ and $BI_{L,C,R}$ predicts nearly identical d' only $BI_{L,R}$ predictions are shown in the upper panel for improved clarity. The lower panel represents predictions from BMFD. The dashed horizontal lines indicate the decision criterion of the models, e.g., differences between test and reference signals resulting in d' values below the criterion are not assumed to be detectable.

The lower part of Table 1 summarizes RMSE and R² between experimental data and predictions for the three model versions. Is it observed that for most binaural experiments the three model versions BMFD, $BI_{L,C,R}$, and $BI_{L,R}$ achieve a comparable prediction performance. Only in experiment 3 (Frequency and interaural phase relationships in wideband conditions) $BI_{L,R}$ achieved a substantially better performance compared to the other two versions. Therefore, it can be stated that $BI_L$ and $BI_R$ are sufficient to explain most of the data of the binaural psychoacoustic experiments used in this study.

Overall, Table 1 showed that the GPSM with binaural BMFD extension, accounts for several monaural and binaural psychoacoustic experiments.

Table 1 about here

## IV. Speech intelligibility evaluation

The binaural model extension was also tested for the headphone-based binaural (dichotic) speech intelligibility experiments of Ewert et al. [2], where SRTs were measured for frontal target speech [German Oldenburger Satztest (OLSA), [72]] in the presence of two co-located or spatially separated maskers with different spectro-temporal characteristics, but identical long-term spectrum.

Four stationary speech-shaped noise (SSN) based maskers, SSN, SAM, BB, and AFS with different spectro-temporal stimulus properties and two speech maskers were used in [2]: The SAM masker was obtained by applying an 8-Hz sinusoidal amplitude modulation with 100% modulation depth to the SSN masker yielding regular temporal modulations coherent across all auditory channels (co-modulation). For the BB masker, the SSN was multiplied with the Hilbert envelope of a broadband speech signal (ten randomly selected OLSA sentences), introducing temporal gaps that reflect the modulations of intact speech. Temporal irregularities of the speech envelope are coherent across all auditory channels. For the across-frequency shifted (AFS) masker, the speech envelope was randomly shifted in eight groups (each consisting of four adjacent auditory frequency channels) resulting in incoherent AMs across auditory channels. As speech maskers, a male version of the International Speech Test Signal (ISTS; [73]), composed of intact continuous speech uttered by six different female talkers in different languages, was used as "nonsense" speech. A single talker (ST) masker used randomly cut parts of ten concatenated OLSA sentences spoken by a different male speaker than in the target OLSA material.

Two spatial target-masker configurations were measured for each masker: In the co-located configuration target and masker sources were placed in front of the receiver (0°). In the spatially separated configuration, the masker positions were changed two both sides at ±60° relative to the frontal direction. Speech intelligibility improvements depending on the spatial separation between target and masker are expressed as SRM. A single masker had a level of 65 dB SPL, and accordingly the presentation of two statistically independent masker sequences resulting in an overall masker level of 68 dB SPL. A detailed description of the experiment can be found in [2].

## A. Results and discussion

Measured and predicted SRTs are represented by gray and black symbols, respectively. Co-located maskers are indicated by closed symbols and separated maskers by open symbols. Predicted SRTs shown in Figure 7 are averaged over 5 repeated simulations each based on 20 OLSA sentences. Each model version was calibrated to the speech material as proposed in [25] by setting the parameters k, q, m, $\sigma_s$ in order to match the SSN data, which are shown in Table 2.

Table 2 about here

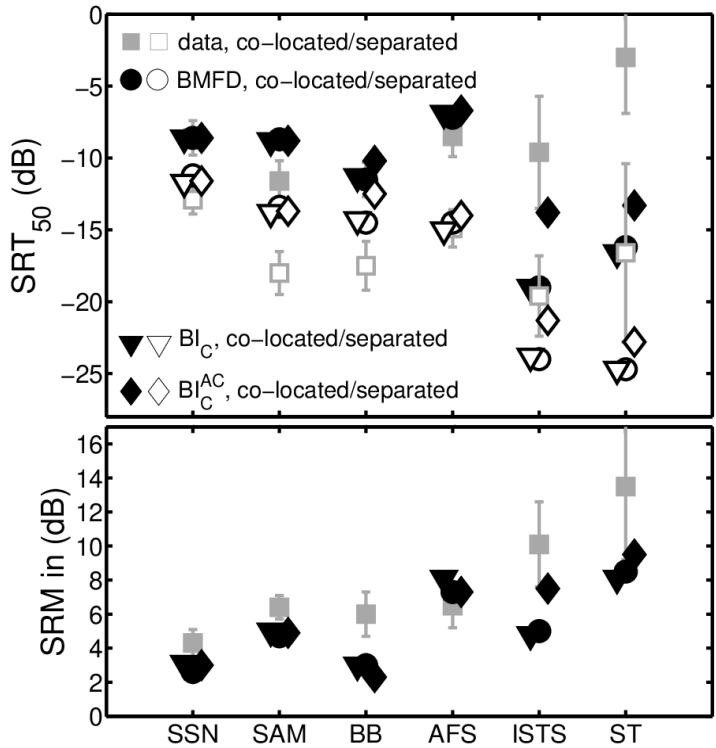

Figure 7: The upper panel shows $SRT_{50}$ results, while the lower panel shows the respective SRM. Data is represented by squares, while predictions are given by circles, triangles, and diamonds, respectively. The spatially co-located (front) and separated masker conditions are indicated by closed and open symbols, respectively.

For noise maskers (SSN, SAM, AFS, and BB) presented co-located to target speech, the highest $SRT_{50}$s were measured for stationary SSN and fluctuating AFS maskers, and listeners took only advantage from listen into dips when speech was presented in fluctuating SAM and BB maskers. The highest $SRT_{50}$ was measured when speech was masked by the single talker (ST), resulting in about 5.5 dB higher thresholds compared to the SSN masker. A spatial separation of target speech and maskers resulted in SRM values ranging between 4.3 and 13.5 dB. The smallest SRM of about 4.3 dB was observed for the SSN maskers, while the largest SRM values of 10.1 and 13.5 dB were observed for ISTS and ST masker.

All model versions were calibrated to account for the co-located SSN masker, while all other thresholds use the same parameters. For co-located predictions based on the BMFD (closed circles in the upper panel of Figure 7) for fluctuating noise maskers BB and AFS largely agree with data, while the predicted $SRT_{50}$ for the SAM maskers is about 3 dB higher than measured $SRT_{50}$. For BMFD the largest differences between predicted and measured SRTs of up to 13 dB can be observed for co-located ISTS and ST maskers. Particularly the ST masker is very similar to the target sentences and makes it difficult for the listener to separate the target from the interfering speech (informational masking, e.g. [74]), which results in high SRTs and high variability across listeners. In contrast to human listeners, the current model, as other intrusive SI models, has a-priori knowledge about the target speech and the masker signals and is only limited by aspects of amplitude modulation and energetic masking (and not informational masking), yielding to substantially lower thresholds for the speech like maskers. For the spatially separated conditions (open circles in the upper panel of Figure 7) BMFD predictions fit well for SSN and AFS while in overestimates the thresholds for SAM and BB and again underestimates thresholds for the speech like maskers ISTS and ST as can be expected (see above). Regarding the SRM (lower panel of Figure 7), BMFD predictions show a good agreement with the data for SSN, SAM (about 2 dB reduced SRM) and AFS. For BB the predicted SRM is about 3 dB lower and for ISTS and ST up to 5 dB lower than the measured SRM. For ISTS and ST these differences are partly caused by larger discrepancies between predicted and measured SRTs in co-located conditions.

In a further step, each of the five BMFD outputs was analyzed to identify the most contributing channel. Here, $BI_C$ with highest sensitivity to the hemispheric midline denoted as $BI_C$ in Figure 7, gave most contribution to SI predictions, that is clearly shown by very similar predictions of BMFD and $BI_C$ in Figure 7. This agrees well with the findings of Ewert et al. [2], where a simple binaural summation of the left and right ear signals (prior to the model) showed similar results for predictions using the binaural speech intelligibility model (BSIM; [12]). For this summed diotic input, BSIM effectively reduces to a similar processing as suggested in the monaural ESII [11] model, using a short-time assessment of power-based

SNRs. In contrast the current BI$_C$ predictions are based on both short-time envelope power and power SNRs. It should be noted that although predictions of both the power pathway of BMFD and BSIM are based on power SNRs, substantial differences exist, like the SNR combination across time frames and auditory channels, which could have an influence on predicted SRTs.

Analyzing the contribution of envelope power and power SNRs, revealed that AM cues are mostly dominant. Predictions only based on envelope power SNRs provided by the center binaural interaction channel are denoted as $BI_C^{AC}$ and shown as diamonds in Figure 7. With exception of the BB masker condition $BI_C^{AC}$-based predictions already explain most of the SRM observed in the data.

Although BI$_C$ does not play an important role for the binaural psychoacoustic experiments in this study, it can successfully account for a large part of the SRM in the speech intelligibility experiments.

## VI. General discussion

The suggested model explores the ability of a strongly simplified, fixed (non-adaptive) binaural interaction stage to account for key aspects of binaural psychoacoustics and speech intelligibility with spatially separated interferers. The investigated 5-channel BMFD stage was incorporated in an existing monaural model using power and envelope power SNR cues. It was demonstrated that the suggested model maintains the ability of the former monaural approach to account for monaural psychoacoustic key phenomena. Binaural psychoacoustics was well covered except for larger discrepancies for time-intensity trading. For speech intelligibility, the key aspects where also predicted with larger discrepancies for speech-like interferers. Here aspects of informational masking which are generally not covered by signal-processing models play a role, as has been previously shown for other speech intelligibility models.

It is conceivable that the current simplified approach might not reach the performance of other "specialist", dedicated monaural and binaural models for psychoacoustics and speech intelligibility for each of the experiments considered here. The value of the current approach is that i) based on former work [9, 16, 19, 20, 26] the suggested model can be assumed to generalize well for other unknown data. This makes the model interesting also in the context of instrumental (spatial) audio quality predictions. ii) Another consideration is that the simple processing in the BMFD stage is generally advantageous for real-time applications, e.g., for control of signal processing algorithms in hearing supportive devices or as hearing aid processing stage itself. iii) The current approach demonstrates that the physiologically motivated hemispheric interaural interaction in mammals (e.g., [42, 43]), as realized here in the two binaural interaction channels $BI_L$ and $BI_R$, is suited to explain a broad variety of perception experiments.

### A. Contribution of binaural interaction and better ear channels

For the binaural psychoacoustic experiments used in this study, the two $BI_L$ and $BI_R$ channels appear sufficient to account for the data. $BI_C$ has only a negligible effect on the predicted data as also indicated by very similar RMSE and $R^2$ values shown in Table I for the model versions including $BI_c$ ($BI_{L,C,R}$) and excluding $BI_c$ ($BI_{L,R}$), except for the binaural experiment 3 on interaural phase effects in wideband conditions: Here predicted thresholds based on $BI_C$ are significantly better than human performance in $N_\pi S_m$ and $N_\pi S_0$ conditions (see middle panels in Figure 6) and accordingly predicted difference pattern for $N_\pi S_m$-$N_0 S_m$ and $N_\pi S_0$-$N_0 S_\pi$ show a large deviation of up to 10 dB at 250 Hz from measured data. In general, both better ear channels $BE_L$ and $BE_R$ did not make any substantial contribution in the binaural psychoacoustic experiments.

For speech intelligibility, the importance of the five BMFD channels is different and $BI_C$ has been shown to account for a large part of the data (see Figure 7). In the current SI conditions, a frontal target was presented in either co-located or spatially separated maskers. In view of the psychoacoustic conditions, the co-located condition can be regarded as $N_0S_0$, while the separated condition can be considered as $S_0$ plus noise with frequency-dependent interaural phase difference. In the separated conditions, the $BI_C$ channel amplifies the coherent frontal target speaker ($S_0$), while spatially separated maskers with IPDs $\neq 0$ are incoherently added or might be partially cancelled.

The role of the five BMFD channels for speech intelligibility can be further assessed by analyzing the distribution of most contributing envelope power and power SNRs across frequency and over the five binaural processing channels (not shown): For all spatially separated conditions, $BI_C$ shows the highest contribution (in agreement with the additive approach in [2]). For the co-located conditions, no large differences in the contributions of all channels are observed. $BI_L$ and $BI_R$ contribute slightly more, resulting in about 1 dB lower SRTs for $BI_L$ and $BI_R$ than for the other three channels. Regarding the SRM, in line with the psychoacoustic experiments, the two better-ear channels contributed less resulting in consistently lower predicted SRM than the three binaural interaction channels. Although $BI_L$ and $BI_R$ might be less important in the current spatial configuration with frontal target where $BI_C$ was most beneficial, they can be assumed to be more important when the target is placed to either side of the head. Moreover, both $BI_L$ and $BI_R$ are also assumed to be important for the evaluation of spatial audio quality as inaccuracies in the audio rendering of sound reproduction systems may alter the spatial properties, e.g., location, apparent source width, of an auditory object.

## B. Comparison of the binaural stage to other literature models

The outputs of the suggested BMFD stage can be considered as a simplification of the delay-gain matrix and the left/right channel in Breebaart et al. [8] or as specific fixed states of the EC model [28]. Given the conceptual similarity of these two models itself and the widespread use of the EC approach as binaural processing stage in numerous auditory models (e.g., [12, 36, 37]), might make the current results interesting for other literature models.

The three $BI_{L,C,R}$ channels are comparable to elements in the matrix of the Breebaart model with according delay and gain in the respective auditory frequency channel. The $BE_{L,R}$ channels are directly comparable to the individual ear signals passed to the detector stage in the Breebaart model, in parallel to outputs of the delay-gain matrix. In the Breebaart model, internal delays up to 5 ms ($\pi$ phase shift at 100 Hz) and a gain difference up to 10 dB between both ears are realized. These parameters broadly cover the current choice in the $BI_{L,R}$ channels. Thus the difference between the suggested model and the Breebaart model is the reduction of degrees of freedom in the binaural interaction stage to parameters that are directly motivated by physiology in mammals.

Similarly, each of the five BMFD outputs represents a specific state of the EC approach. Again the difference is that the EC stage can realize arbitrary delays and gains (for the equalization of the noise in the left and right channel) to optimally cancel the noise at the output, while $BI_{L,C,R}$ represents a fixed, potentially suboptimal, realization of the EC process. Alternatively, the left or right ear input can be directly routed to the EC output, comparable to the better-ear channels $BE_{L,R}$ in the current BMFD stage.

Based on the five BMFD outputs, envelope power and power SNRs are calculated and combined to give an overall $d'$. In contrast to other models like the B-sEPSM [37] and BSIM [12] where SI prediction are either based on envelope power SNRs or power SNRs, this approach combines both types of SNRs. As shown in Figure 7, envelope power SNRs capture most of the measured SRM. It should be noted that predictions only based on power SNRs also agree with the measured SRM pattern, but tend to overestimate measured SRM. For fluctuating maskers, SRTs predicted by power SNRs are often substantially lower than measured SRTs, which was also observed in Biberger and Ewert [26]. As suggested in [26], a forward masking function or SNR limitation could be applied to counteract that effect.

The envelope power $SNR_{envW,i}(p,n)$ and $SNR_{DC,j}(p)$ are combined across the five BMFD outputs by taking the largest value for each time frame within each auditory and modulation channel. Such a procedure allows fast switching between the five BMFD outputs, in line with findings of Siveke et al. [75]. However, psychophysical studies (e.g., [70], also considered here, see Fig 5.) and a recent SI study of Hauth and Brand [76] implied some limitations of the binaural auditory system in following temporal changes of ITDs (or IPDs). This is often referred to as binaural sluggishness, and suggests binaural temporal windows with time constants of up to about 200 ms. The current model has the same time constants for monaural and binaural interaction channels, resulting in the same slope of the transition in the data of Kollmeier and Gilky [70], see Figure 5. Thus, for some conditions prediction performance could be improved when aspects of (task dependent) binaural sluggishness are integrated into the suggested model by using a temporal window as suggested in [8].

### C. Model limitations and simplification of physiological processes

The current L-R and R-L processing after delay and amplification in the current $BI_{L,R}$ channels represents a strongly simplified realization of hemispheric processing as suggested in more detailed models (e.g., [41, 77]) based on (simulated) neuronal responses. A key feature of these approaches is the characteristic (hemispheric) net neural activation as a function of ITD for high frequencies in the lateral superior olive (LSO) and for low frequencies in the medial superior olive (MSO), see, e.g., bottom row in Figure 5 of [41].

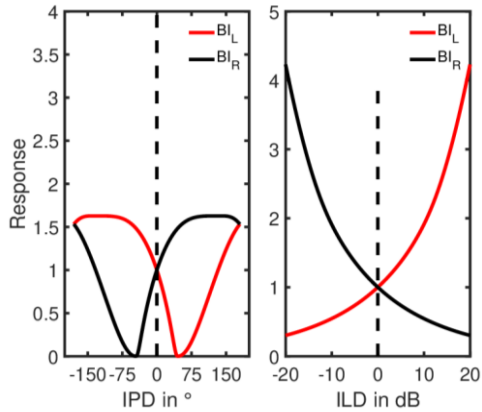

Figure 8: Response of the $BI_L$ and $BI_R$ channels as a function of IPD (left panel) and ILD (right panel) for a 500 Hz pure tone. Negative IPDs indicate left ear leading, while negative ILDs indicate right ear more intense. Note that for clarity, amplitude and phase jitter were turned off.

The (hemispheric) net neural activation is only partly resembled with the current subtraction process of the half-wave rectified continuous time signal as illustrated in Figure 8 and is reminiscent of to that observed in the LSO (first two rows in Figure 5 of [41]). The left panel of Figure 8 shows the linear response of $BI_L$ (red lines) and $BI_R$ (black lines), normalized to the response at 0° IPD, as a function of the IPD (negative sign indicates left ear leading, no ILD) for $\tau$ (delay) of $\pi/4$ and $\alpha$ of 3. The strongest contralateral inhibition occurs when the contralateral ear is leading with an IPD of $\tau$. The least inhibition occurs when the ipsilateral ear is leading with an IPD of $\pi-\tau$, resulting in internal phase differences of $\pi$ between the excitatory and inhibitory channels. The current $\tau$ value of $\pi/4$ provides a sufficient steep slope around zero IPD to ensure a sufficient sensitivity for small interaural phase differences and is in line with physiological findings. Smaller values would further increase IPD sensitivity and would improve predictions for data of the ITD experiment shown in Figure 2. The $\alpha$ factor of 3 was selected empirically and leads to a complete inhibition by a contralateral leading ear with an up to 10 dB lower level. Larger values would widen the troughs in the response pattern in the left panel of Figure 8, while smaller values would result in narrower troughs. α values ranging between 3 and 5 resulted in similar prediction performance. The current α agrees well with range of interaural gain differences applied in the Breebaart model. The right panel of Figure 8 represents the linear response as a function of the ILD (negative sign indicates right ear more intense, no IPD). The response of the ipsilateral ear increases as the ipsilateral ear is more intense while inhibition occurs for contralaterally more intense sounds.

In more detailed neural model assumptions (e.g., [41, 42]), the hypothesis of timed inhibition is that the contralateral inhibitory post-synaptic potential (IPSPcontra) precedes the contralateral excitatory PSP for low-frequency processing in the MSO, resulting in a delay of the contralaterally evoked net excitation and the observed hemispheric excitation as a function of ITD. The delayed excitatory interaction, as well as the temporal smearing of excitatory and inhibitory effects represented in the PSPs are not covered by the current (over) simplified model. Moreover, different processing in the LSO and MSO for low and high frequencies, respectively, is observed in the physiology. Conversely, the current model only uses subtraction of the waveforms, disregarding details of PSP simulation, resembling (envelope) ITD processing assumed in the LSO for high frequencies (see center panel of Figure 5 in [41]). This inhibitory processing is used for all frequencies, involving interaural temporal fine structure (TFS) differences at low frequencies and temporal envelope differences at high frequencies. An improvement of the current model can be expected when incorporating both excitatory and inhibitory effects more faithfully, however, at the cost of simplicity.

To compare inhibitory vs excitatory interaction in the context of the current model, we replaced the current subtractive (inhibitory) processing by an additive (excitatory) processing, resulting in an overall similar prediction performance for the psychoacoustic experiments. However, large $\tau$ values above about $3\pi/4$ had to be used to ensure sufficiently large response differences between stimuli with and without interaural phase shifts. Although, the additive processing also explained most of data from the binaural psychoacoustic experiments used in this study, the SRM predictions in SI experiments were often substantially lower than observed in data. Accordingly, the RMSE between predicted and measured SRM was higher for the additive processing (RMSE of 5.5 dB) than for the current subtractive processing (RMSE of 3.3 dB).

### D. Relation to binaural signal processing algorithms

The five outputs of the suggested fixed BMFD stage can be translated to binaural signal processing, potentially applicable in hearing supportive devices. The difference between the model stage and audio signal processing is that the model operates on a half-wave rectified internal representation, whereas audio signal processing operates on the input waveform at the ears. This difference is important for the binaural interaction channels where the ear signals are combined after nonlinear processing in the model. As outlined in the introduction, the processing of $BI_C$ was designed to resemble the effect of summation of the waveform in the ears. For $BI_L$ and $BI_R$, the subtraction of the unipolar (half-wave rectified) signals is followed by a maximum operation with zero, which makes the result more comparable to a subtraction of the waveforms. Thus, as a signal processing algorithm, $BI_C$ represents a (spatially broadly tuned) fixed broadside beamformer (tuning to front and back). Taking the phase delays and subtraction into account, $BI_L$ and $BI_R$ conceptually represent fixed (non-adaptive) first-order differential microphone beamformers with a (frequency-dependent) steering vector. Finally, taking the head shadow effect into account, $BE_L$ and $BE_R$ can be interpreted as beamformers pointing to the left and the right. Thus, the BMFD in the current model suggest that the auditory system selects the favorable output of five beamformers in time-frequency frames, depending on the task and spatial configuration of the input.

In comparison to the adaptive EC model, the current approach cannot optimize parameters to specifically cancel certain signal parts (or directions) as in the adaptive differential microphone. Further simplifying the current selection of the optimal BMFD channel in time-frequency frames to the selection of a single broadband channel, the BMFD might be applicable in hearing aid processing as five spatially broadly tuned binaural beamformers from which the optimal output is selected, e.g., based on direction of arrival of the intended target. Such simplistic beamformers might also be better suited in ecologically valid situations with head movements where the additional benefit of more elaborated processing might be limited (e.g., [78]). Indicated by the current speech intelligibility results for a frontal (speech) target, humans appear to just use a simple broadside binaural beamformer ($BI_C$).

## VII.  Summary and conclusions

The main goal of this study was to examine how well a modelling approach with strongly simplified assumptions about a fixed (non-adaptive) binaural interaction processing can predict data from both binaural psychoacoustic and speech intelligibility experiments. For this, the generalized power spectrum model [26] was extended by a five channel binaural matrix feature decoder, comprising two better-ear and three binaural interaction channels, to account for monaural and binaural aspects in psychoacoustic and speech intelligibility experiments. The binaural processing comprises the left (L) and right (R) better ear channels, the L+R channel ($BI_C$) and two L-R ($BI_L$) and R-L ($BI_R$) channels incorporating a fixed phase delay ($\pi/4$). The model was tested in a monaural and binaural "benchmark" of overall 13 psychoacoustic experiments and 6 conditions of a speech intelligibility experiment from literature. The following conclusion can be drawn:
- The suggested binaural model accounts for several temporal and spectral key aspects in classical binaural experiments from literature and also explains a large amount of spatial release from masking in speech intelligibility experiments. The model maintains the predictive power of the earlier monaural approach for monaural psychoacoustics.

- In the psychoacoustic experiments of this study, the L-R and R-L binaural interaction channels, physiologically motivated by hemispheric processing, were most important as the target signal often contained an interaural phase shift ($S_\pi$). The L+R "midline" channel played no important role.

- For the current speech intelligibility predictions, with a frontal target and spatially separated maskers (somewhat similar to a $S_0$ plus noise with frequency-dependent interaural phase difference condition in psychoacoustics), the L+R channel was most important to account for SRT and the spatial release from masking.

- Overall, the results show that human performance in binaural task might be based on a smart selection of spectro-temporal segments at the output of only a few fixed binaural interaction channels.

## VIII. ACKNOWLEDGMENTS

We would like to thank M. Dietz, B. Eurich, and J. Encke for helpful remarks. We would also like to thank the members of the Medizinische Physik and Birger Kollmeier for continued support. This work was supported by the Deutsche Forschungsgemeinschaft (DFG – 352015383 – SFB1330 A2 and DFG – 390895286 – EXC 2177/1).

# X. Tables

Table 1: Root-mean square errors (RMSE) and coefficient of determination ($R^2$; squared cross-correlation coefficient) between data and model predictions for the monaural and binaural psychoacoustic experiments.

| Monaural Experiments | | | | | | |
|---|---|---|---|---|---|---|
| Experiments | BMFD | | $BI_{L,R}$ | | mr-GPSM [26] | |
| | RMSE | $R^2$ | RMSE | $R^2$ | RMSE | $R^2$ |
| 1. Hearing threshold | 3.3 dB | 0.99 | 3.3 | 0.99 | 1.7 dB | 0.99 |
| 2. Intensity JNDs | 0.2 dB | 0.66 | 0.2 | 0.64 | 0.3 dB | 0.57 |
| 3. Tone in noise | 1.3 dB | 0.99 | 1.3 | 0.99 | 2.1 dB | 0.99 |
| 4. Spectral masking | 9.5 dB | 0.82 | 9.5 | 0.8 | 7.9 dB | 0.9 |
| 5. AM detection | 4.0 dB | 0.71 | 4 | 0.78 | 4.5 dB | 0.68 |
| 6. AM discrimination | 2.4 dB | 0.94 | 2.4 | 0.92 | 1.6 dB | 0.94 |
| 7 AM masking | 4.6 dB | 0.77 | 4.7 | 0.79 | 6.2 dB | 0.73 |
| Binaural Experiments | | | | | | |
| Experiments | BMFD | | $BI_{L,C,R}$ | | $BI_{L,R}$ | |
| | RMSE | $R^2$ | RMSE | $R^2$ | RMSE | $R^2$ |
| 1. ITD discrimination | 0.019 ms | 0.89 | 0.019 | 0.9 | 0.019 ms | 0.93 |
| 2. IID discrimination | 0.5 dB | 0.002 | 0.5 | 0.0014 | 0.5 dB | 0.005 |
| 3. Frequency and interaural phase relationships in wideband conditions | 9.1 dB | 0.86 | 8.5 dB | 0.85 | 6.7 | 0.88 |
| 4. $N_0S_\pi$ depending on signal duration | 2.9 dB | 0.92 | 3.0 | 0.92 | 3.2 dB | 0.9 |
| 5. Temporal phase transition | 2.6 dB | 0.80 | 2.7 dB | 0.8 | 2.7 dB | 0.81 |
| 6. Time-intensity-trading | 0.5 | 0.38 | 0.6 | 0.58 | 0.6 | 0.61 |

Table 2: Parameter settings of the three model versions to match the co-located SSN data. The $\bar{k}$ value results from averaging the individual k values from five repeated simulations.

|  | $\bar{k}$ | q | m | $\sigma_s$ |
|---|---|---|---|---|
| BMFD | 0.6 | 0.5 | 50 | 0.6 |
| $BI_C$ | 0.72 | 0.5 | 50 | 0.6 |
| $BI_C^{AC}$ | 0.72 | 0.5 | 50 | 0.6 |

# XI. Figure captions

Figure 1: Block diagram of the GPSM with BMFD extension. After peripheral processing, the left and right ear signals are binaurally processed by using the BMFD that provides two better-ear channels $BE_L$ and $BE_R$ and three binaural interaction channels $BI_L$, $BI_C$, $BI_R$. For each of the five BMFD outputs, envelope power and power SNRs are calculated on short-time frames and then combined across the five channels of the BMFD and across auditory and modulation channels, resulting in a sensitivity index $d'_{env}$ based on envelope power SNRs and $d'_{DC}$ based on power SNRs. The final combined $d'$ was then compared to a threshold criterion that assumes that a signal is detected if $d' > (0.5)^{1/2}$.

Figure 2: empirical data (filled symbols) and model predictions (open symbols) for ITD thresholds in ms (upper panel) and IID thresholds in dB (lower panel).

Figure 3: Empirical data (filled symbols) and model predictions (open symbols) for masked thresholds for wideband $N_0S_m$ (upper-left panel), $N_0S_\pi$ (upper-right panel), $N_\pi S_m$ (middle-left panel), and $N_\pi S_0$ (middle-right panel) conditions as a function of the frequency of the signal. Differences in thresholds between the $N_\pi N_m$ and $N_0S_m$ are shown in the lower-left panel, while the lower-right panel represents differences in threshold between $N_\pi S_0$ and $N_0S_\pi$.

Figure 4: Empirical data (filled symbols) and model predictions (open symbols) for $N_0S_\pi$ thresholds as a function of the signal duration. Data and predictions are shown for signal frequencies of 500 Hz (left panel) and 4 kHz (right panel).

Figure 5: Empirical data (filled symbols) and model predictions (open symbols) for $N_\pi N_0 S_\pi$ (upper-left panel) and $N_\pi N_0 S_\pi$ (upper-right panel) thresholds as a function of the temporal position of the signal center relative to the masker-phase transition. Monaural thresholds for $N_\pi N_{\pi,-15dB} S_\pi$ and $N_{\pi,-15dB} N_\pi S_\pi$ are shown in the lower-left and lower-right panels. Filled symbols represent four subjects measured by Kollmeier and Gilky [70].

Figure 6: Empirical data (grey lines, closed symbols) and model predictions (black lines, open symbols) for the time-intensity trading experiment of Hafter and Carrier [71] with different ITDs of 0, 10, 20, 30, and 40 μs. The ordinate represents $d'$, while the abscissa represents the ILD in dB. Since $BI_{L,R}$ and $BI_{L,C,R}$ predicts nearly identical $d'$ only $BI_{L,R}$ predictions are shown in the upper panel for improved clarity. The lower panel represents predictions from BMFD. The dashed horizontal lines indicate the decision criterion of the models, e.g. differences between test and reference signals resulting in $d'$ values below the criterion are not assumed to be detectable.

Figure 7: The upper panel shows SRT$_{50}$ results, while the lower panel shows the respective SRM. Data is represented by squares, while predictions are given by circles, triangles, and diamonds, respectively. The spatially co-located (front) and separated masker conditions are indicated by closed and open symbols, respectively.

Figure 8: Response of the BI$_L$ and BI$_R$ channels as a function of IPD (left panel) and ILD (right panel) for a 500 Hz pure tone. Negative IPDs indicate left ear leading, while negative ILDs indicate right ear more intense. Response shown in both panels are based on the same τ and α values of $\pi/4$ and 3 as they were used for all simulations in this study. Note that for clarity, amplitude and phase jitter were turned off.